\documentclass[final,1p,times]{article}

\usepackage{amssymb}

\usepackage{graphicx}

\usepackage{epsfig}

\usepackage{amssymb}
\usepackage{amsmath}

\begin{document}


\title{Stability conditions for fermionic Ising spin-glass models in the presence of a transverse field}

\author{S. G. Magalh\~aes$^1$, F. M. Zimmer$^2$ and C. V. Morais$^1$
\\ \\
{\it
$^1$Universidade Federal de Santa Maria 97105-900},\\ {\it Santa Maria, RS, Brazil}
\\
{\it $^2$Universidade do Estado de Santa Catarina, 89223-100,}
\\
{\it Joinville, SC, Brazil}}

\date{}
\maketitle
\begin{abstract}

The stability of spin-glass (SG) phase is analyzed in detail for a fermionic Ising SG (FISG) model in the presence of a magnetic transverse field $\Gamma$.
The fermionic path integral formalism, replica method and static approach 
have been used to obtain the thermodynamic potential within one step replica symmetry breaking ansatz.  
The replica symmetry (RS) results show that the SG phase is always unstable against the replicon. 
Moreover, the two other eigenvalues $\lambda_{\pm}$ of the Hessian matrix (related to the diagonal elements of the replica matrix) 
can indicate an additional instability to the SG phase, which enhances when $\Gamma$ is increased.
Therefore, this result suggests that the study of the replicon can not be enough to 
guarantee the RS stability in the present quantum FISG model, especially near the quantum critical point. 
In particular, the FISG model allows  changing the occupation number of sites, so  
one can get a first order
transition when the chemical potential exceeds a certain value.
In this region, the replicon and the $\lambda_{\pm}$ indicate instability problems for the SG solution  close to all range of first order boundary.

\end{abstract}

\section{Introduction}

The interplay between disorder, frustration and quantum effects
presents several 
challenging issues.
For instance, 
the
infinite-range Ising spin-glass  model
in a transverse magnetic field 
has been studied with 
various
techniques that  
show controversial
results. 
It is well known that 
the classical Sherrington-Kirkpatrick (SK) model \cite{sherrington} presents a continuous 
phase transition in which the spin-glass (SG) phase has the free energy landscape composed by
many almost degenerated thermodynamic states separated by infinitely high barriers \cite{Fischer}.
The controversy 
is 
whether or not the 
quantum tunneling between free energy barriers  activated by the transverse field is able 
to restore the replica symmetry (RS) in the infinite-range Ising spin-glass model. \cite{thirumalai,usadel,gold,koreanos,eduardo,zimmerprb}.
To answer this question, the usual procedure
is to investigate the behavior 
of the so-called
replicon (the transversal eigenvalue of the Hessian matrix) 
to
check 
the local  stability of 
RS solution. 
No much attention is 
given to the longitudinal eigenvalues since, as in the classical case,
they would not give
any additional important information in 
such  an analysis 
\cite{thirumalai,koreanos}. 
Nevertheless, 
there is, at least, 
one 
formulation of 
the infinite-range Ising spin-glass problem, the fermionic Ising spin-glass 
(FISG) model \cite{albagusmao,wener,oppermannmuller}, in which 
the role of  the quantum effects on 
the longitudinal eigenvalues could also be 
important. 

The FISG model is defined by 
spin operators which are represented by bilinear
combinations of fermion operators of creation and destruction. That makes 
this
model 
a quite 
adequate
framework
to study 
the competition between spin-glass and, for instance, Kondo effect or superconductivity.
\cite{magal1,feldmann,magal5}.
Those spin operators
act on Fock space with four states per site, two of them nonmagnetic. 
In fact,
this model 
can also be
formulated in two versions \cite{Alba2002,zimmermagal}. 
In the 
first one 
($2S$ model),  it is used a restriction 
in the number operator 
that eliminates
the contribution of nonmagnetic states. 
That would be
equivalent to study the problem in the spin space. 
In the 
second version 
($4S$ model), magnetic 
and nonmagnetic
states are admitted.
In this 
case, 
an important aspect
is 
the connection between 
spin and charge correlations 
\cite{oppermann}. That can be seen clearly, for instance,
in the relationship between 
$n$ (the average of occupation of fermions per site)
and 
the diagonal replica matrix elements \cite{feldmann,oppermann}.
Therefore, in the $4S$ model,  
variations of charge occupation 
can 
influence the magnetic properties 
 not only leading the onset of the spin-glass phase, 
 but 
also  
changing the
nature of the phase boundary. 
For instance, the phase diagram of the 
$4S$ 
model presents a tricritical point 
\cite{oppermann} and a reentrance in the first order boundary phase for specific values of the chemical potential $\mu$ \cite{magal5}.
The mentioned connection 
can also appear in the behavior of longitudinal eigenvalues of the Hessian matrix.
Indeed, 
in the $4S$ model 
these eigenvalues
present a non-trivial behavior, 
in which they become complex \cite{feldmann,oppermann}.

The 
presence of a transverse field $\Gamma$
produces
important changes 
in the 
phase diagram
of $2S$ and $4S$
models. 
For instance, the increase of $\Gamma$ leads the freezing temperature 
($T_{f}$) of both models
to a 
Quantum Critical Point ($QCP$) at $\Gamma=\Gamma_{c}$ \cite{zimmerprb,Alba2002}.
Particularly, 
for the $4S$-model, when $\mu\neq 0$,
the increase of $\Gamma$ not only 
depresses the second order part of 
$T_{f}$
and  
the position of the tricritical point 
 but also 
destroys 
the reentrance in the first order part of $T_{f}$ \cite{magal5}. 
In other words,
$\Gamma$ induces quantum spin flipping mechanisms that  affect  the replica matrix 
elements  strongly, 
in particular,  the diagonal ones.
Because of that,
$\Gamma$
could
interfere in the replicon ($\lambda_{AT}$) and,  it should be remarked,  
in the 
longitudinal
eigenvalues ($\lambda_{\pm}$) of Hessian matrix for both $2S$ and $4S$ models.

The previous discussion leads to 
the question: 
how 
exactly 
 would $\Gamma$
affect
$\lambda_{\pm}$ in these two models?
In the $2S$ model,  the diagonal replica matrix elements become dependent on temperature when $\Gamma>0$, which could reflect on the stability of the SG phase. Thus, it can be a mechanism to change the behavior of $\lambda_{\pm}$
since,
at $\Gamma=0$, 
this
model reproduces 
the basic features of the classical SK  one.
 One can also remark that
$\Gamma$
in the $2S$ model
produces effects which are 
limited
to the magnetic properties.
Nevertheless, that is not 
the case 
for the 
$4S$ one.
In this model,
 there is a mixing between spin and charge correlation functions 
which is important to determine  not only magnetic properties \cite{magal5}  but also charge ones \cite{feldmann}.
 In this particular model,
even for $\Gamma=0$, 
the diagonal replica matrix elements have a relevant role for the SG solution when $\mu$ is increased, which can lead to a non-trivial behavior of $\lambda_{\pm}$, as discussed 
previously.
Consequently,
one can expect that the combined effects of variations of $\Gamma$ and $\mu$ could enhance 
the non-trivial behavior 
of $\lambda_{\pm}$.
Actually, the variations of these independent parameters create a complicated interplay between charge and spin degrees of freedom in which 
 both magnetic properties and charge distribution, for instance, are affected.
 However, such 
redistribuition  of charge
can 
influence the magnetic correlation functions themselves. 
As a result, the behavior of $\lambda_{\pm}$ in the $4S$ model 
can be considerably more complicated than the $2S$ one.

Therefore, the main purpose of this paper is to 
study
in detail the 
RS 
stability 
for both the $2S$ and $4S$ models in the presence of a transverse magnetic field $\Gamma.$ 
In particular, for the $4S$ model, 
special attention is given to investigate effects of combined variations 
from $\Gamma$ 
and
$\mu$ 
on $\lambda_{AT}$ and $\lambda_{\pm}$. Furthermore, 
it is also obtained the behavior of $n$ inside the
SG 
phase
to clarify  
how 
spin and charge correlation functions influence each other.  In order to accomplish these goals,  the partition function is obtained in the path integral formalism in which the spin operators are given as bilinear combinations of Grassmann fields. The thermodynamic potential is found within the static approach ($SA$)  and the one step replica symmetry breaking ($1S\mbox{-}RSB$) scheme, which allows to derive $n$. In particular, to obtain the $2S$ model, it is introduced the same restraint used in Refs. \cite{wener,Alba2002} to avoid the two nonmagnetic eigenstates. Finally, the local stability of the $RS$ solution is studied by obtaining $\lambda_{AT}$ and $\lambda_{\pm}$. 

The present work the static approach 
which neglects time fluctuations of spin-spin 
correlation functions  is used \cite{braymoore}.  
It is clear that the $SA$ is unable to provide reliable quantitative results for the thermodynamics at very low temperature \cite{theo}. However, it
has  also been 
shown by Grempel and Rozemberg \cite{grampel} that static ansatz gives reliable results when the
temperature is not too low.  In fact, the results found in Ref. \cite {grampel} for the spin-spin correlation function 
$Q(\tau)$ indicate a behavior which is approximate to the same of  their classical counterpart within an 
interval of temperature around  a continuous $SG$ transition. 
 On the other hand, when a first order transition is present, it would be expected that $SA$ could produce naturally reasonable results around the transition \cite{theo}. This interval of reliability for both kind of transition justifies the use of $SA$ for purposes of this work.

It is important to mention that the scenario described for the 
$4S$-model when $\Gamma=0$ has been already found in another model. 
The classical Ghatak-Sherrington (GS) model \cite{ghatak} 
is a particularly interesting example which displays a spin-glass first order 
phase 
transition \cite{crisantileuzzi,mottishawsherrington,almeidalage,franciscosalinas}. 
In the 
GS model,
there are three distinct eigenvalues of the Hessian matrix
in the limit of number of replicas $n\rightarrow 0$. 
As in the classical SK model \cite{almeida}, the 
replicon ($\lambda_{AT}$) 
is negative in the whole SG region. 
However, 
other two eigenvalues ($\lambda_{\pm}$)
can be
complex or negative \cite{mottishawsherrington,almeidalage,franciscosalinas}. 
Actually, there is a close relationship between GS and FISG models \cite{castillo}.
In particular, the thermodynamics
described 
in the GS model 
can also be found in 
the 
$4S$-model, such as
the 
reentrance in the first order boundary phase. 
This similarity of the thermodynamics between these two models
is obtained by a mapping of their spin-glass order parameters equations, within the $SA$ and RS solution, 
given
by a relationship between the chemical potential $\mu$ and 
the anisotropic constant $D$  of the GS model
\cite{feldmann}. In that sense, the GS model 
can be a reference  
to check the correctness of results obtained in the 
$4S$-model when $\Gamma\rightarrow 0$.

The organization of this article is as follows: in section \ref{formulation}, the FISG model in a $\Gamma$ field is presented. The thermodynamic potential is obtained within the SA ansatz and 
$1S\mbox{-}RSB$ solution. In this section, it is also done a detailed study of 
the local stability of the $RS$ solution for the present FISG model. Section \ref{results} is devoted to describe the numerical results. Section \ref{conclusions} is left for conclusions.

\section{General Formulation}
\label{formulation}

The infinite-range Ising spin-glass model in the presence of 
a transverse magnetic field is described by:
\begin{equation}
\hat{ H}= -\sum_{i\neq j} J_{ij}\hat{S}_{i}^{z} \hat{S}_{j}^{z} 
-2 \Gamma \sum_{i} \hat{S}_{i}^{x}\label{ham}
\end{equation}
\noindent
where the sums are run over the $N$ sites of a lattice. The exchange interaction 
$J_{ij}$ among all pairs of spins is assumed to be a random variable with a Gaussian
distribution \begin{equation}
 P(J_{ij})=\sqrt{N/(32 \pi J^{2})}\exp(-J_{ij}^2 N/32 J^{2}).
\label{distrib}
\end{equation}
The spin operators are defined as \cite{Alba2002}:
\begin{equation}
\hat S_{i}^{z}=\frac{1}{2}[\hat{n}_{i\uparrow}-\hat{n}_{i\downarrow}]~,
~~~~~~
\hat S_{i}^{x}=\frac{1}{2}[c_{i\uparrow}^{\dagger}c_{i\downarrow}
+c_{i\downarrow}^{\dagger}c_{i\uparrow}]
\label{ope}
\end{equation}
\noindent
where $\hat{n}_{i\sigma}=c_{i\sigma}^{\dagger}c_{i\sigma}$ is the 
number operator, $c_{i\sigma}^{\dagger}~(c_{i\sigma})$ is fermion 
creation (destruction) operator and $\sigma=\uparrow$ or $\downarrow$ 
indicate the spin projections.

The Hamiltonian given in Eq. (\ref{ham}) is defined on the Fock space where there are four states per site:
one state with no fermion, two states with a single fermion and one state with two fermions. 
Consequently, there are two nonmagnetics states.
This work considers two models: the $4S$ model that allows the four possible states per site and 
the $2S$ model, which restricts the spin operators to act 
on a space where the nonmagnetic states are forbidden. 
Therefore, the $2S$ model requires a restriction to 
remove the contribution of these nonmagnetic states.  
It can be obtained by computing only sites occupied by one fermion ($n_{i\uparrow}+n_{i\downarrow}=1$  
at every site) in the partition function trace \cite{wener,Alba2002}.

In this fermionic problem the partition function is expressed 
by use of the Lagrangian path integral formalism in terms of anticommuting Grassmann fields  ($\phi$ and $\phi^*$)
\cite{Alba2002}.
The restriction in the 2S-model is obtained by using the Kronecker delta function \cite{wener,Alba2002}. 
Therefore, adopting an integral representation for this delta function, one can express the partition function for both models in a compact form:
\begin{equation}
Z\{y\}=\mbox{e}^{\frac{s-2}{2}N\beta\mu}\int D(\phi^* \phi)\prod_{j}^{}\frac{1}{2\pi}\int_{0}^{2\pi}dx_j\mbox{e}^{-y_j}\mbox{e}^{A\{y\}}
\end{equation}
where
\begin{equation}
A\{y\}=\int_{0}^{\beta}d\tau\left\{\sum_{j,\sigma}\phi_{j\sigma}^{*}(\tau)\left[
 \frac{\partial}{\partial \tau} +\frac{y_j}{\beta}\right]\phi_{j\sigma}(\tau) - H\left(\phi_{j\sigma}^{*}
 (\tau),\phi_{j\sigma}(\tau)\right) \right\}
\label{action},
\end{equation}
$\beta=1/T$ ($T$ is the temperature), $y_{j}=ix_j$ for the $2S$-model or $y_{j}=\beta\mu$ for the $4S$-model, 
$\mu$ is the chemical potential and $s$ is the state number per site allowed in each model.

Now, it follows the standard procedures in which
the replica method \cite{Fischer},
\begin{equation}
\beta \Omega=-\frac{1}{N}\langle \ln Z\{y\}\rangle_{J_{ij}}=-\frac{1}{N}\lim_{n\longrightarrow 0}\frac{\langle Z\{y\}^n \rangle_{J_{ij}} - 1}{n}
\label{omega}
\end{equation}
is used to get the configurational averaged thermodynamic potential. The replicated partition function  
$\langle Z\{y\}^n \rangle_{J_{ij}}$ is then given by:
\begin{equation}
\langle Z\{y\}^n\rangle_{J_{ij}}=\mbox{e}^{\frac{s-2}{2}N\beta\mu}{\cal N}\int_{-\infty}^{\infty} \prod_{\alpha,\gamma}^{n} dq_{\alpha\gamma} 
\int_{-\infty}^{\infty} \prod_{\alpha}^{n} dp_{\alpha}\exp\left[N \Omega_{n}(q_{\alpha\gamma},
p_{\alpha})\right]
\end{equation}
where ${\cal N}=(\beta J\sqrt{N/2\pi})^{n(n+1)/2}$ and $\alpha=1,2,\cdots,n$ is the replica index.  Within the static approximation, one has:
\begin{equation}
\Omega_{n}(q_{\alpha\gamma},p_{\alpha})=-{\beta^2J^2}\sum_{(\alpha,\gamma)}q_{\alpha,\gamma}^{2}
-\frac{\beta^2J^2}{2}\sum_{\alpha}p_{\alpha}^{2}+ \ln\Lambda\{y\}
\label{stacionario}
\end{equation}
where the Fourier representation is used to express:
\begin{equation}
\Lambda\{y\}=\prod_{\alpha}\frac{1}{2\pi}\int_{0}^{2\pi}dx_{\alpha}\mbox{e}^{-y_{\alpha}}
\int D[\phi_{\alpha}^{*},\phi_{\alpha}]\exp[H_{eff}],
\label{efetivo}
\end{equation}
\begin{equation}
H_{eff}=\sum_{\alpha} A_{0\Gamma}^{\alpha} + 4\beta^{2}J^{2}\left( 
\sum_{\alpha}p_{\alpha}S_{\alpha}^{z}S_{\alpha}^{z}
+2\sum_{(\alpha,\gamma)}q_{\alpha\gamma}S_{\alpha}^{z}S_{\gamma}^{z} \right)  
\label{heff}
\end{equation}
with the convention that $(\alpha,\gamma)$ indicates a distinct pair of replicas and the definitions:
\begin{equation} 
A_{0\Gamma}^{\alpha}=\sum_{\omega}\underline{\varphi}_{\alpha}^{\dagger}(\omega)(i\omega+y_\alpha+
\beta\underline{\sigma}^{x})
\underline{\varphi}_{\alpha}(\omega), ~~S_{\alpha}^{z}=\frac{1}{2}\sum_{\omega}
\underline{\varphi}_\alpha(\omega)\underline{\sigma}^{z}\underline{\varphi}_\alpha(\omega),
\end{equation}
$\underline{\varphi}_{\alpha}^{\dagger}(\omega)=\left(\phi_{\uparrow\alpha}^{*}(\omega)
~~\phi_{\downarrow\alpha}^{*}(\omega)\right)$,
$\omega=\pm \pi,\pm 3\pi,\cdots $.  $\underline{\sigma}^x$ and $\underline{\sigma}^z$ are the Pauli matrices.

Therefore, in the 
$1S\mbox{-}RSB$ Parisi scheme \cite{parisi}, 
the thermodynamic potential is written from Eq. (\ref{omega}) as
\begin{equation}
\begin{split}
\beta \Omega&=\frac{(\beta J)^{2}}{2}[(m-1)q_{1}^{2}-m q_{0}^{2}+p^{2}]-\frac{(s-2)}{2}\beta\mu \\ &-\frac{1}{m}\int Dz\ln \left\lbrace \int Dv [K
(z,v)
]^{m} \right\rbrace -\ln 2
\label{grandpot}
\end{split}
\end{equation}
where 
\begin{equation}
K
(z,v)
=\frac{(s-2)}{2}\cosh(\beta\mu)+\int D\xi \cosh[\sqrt{\Delta
(z,v,\xi)
}],
\end{equation}
with $\Delta
(z,v,\xi)
=[\beta h
(z,v,\xi)]^{2}
+(\beta\Gamma)^{2}$ and
\begin{equation}
 h
(z,v,\xi)
=J\sqrt{2}(\sqrt{q_{0}}z+\sqrt{q_{1}-q_{0}}v+\sqrt{p-q_{1}}\xi),
 \label{hfield}
\end{equation}
where $Dx=dx$e$^{-x^2/2}/\sqrt{2\pi}$ ($x=z,~v$ or $\xi$).
In Eqs. (\ref{grandpot}) and (\ref{hfield}), $q_0$ and $q_1$ are the $1S\mbox{-}RSB$ order parameters and $p=q_{\alpha\alpha}=\langle S^{z}_{\alpha}S^{z}_{\alpha}\rangle$ is the diagonal replica spin-spin correlation that is related to the diagonal replica matrix elements.  The parameters $q_0$, $q_1$, $p$ and $m$ are given by the extreme condition of the 
thermodynamic potential Eq. (\ref{grandpot}).
The RS solution is recovered when $q_0=q_1 (\equiv q)$ and $m=0$.

Particularly, in the 4S model, the chemical potential $\mu$ controls  
the average occupation of fermions per site 
$n=\langle n_{\uparrow}+n_{\downarrow}\rangle$ 
 which is obtained  as: 
\begin{equation}
n=1+\tanh(\beta\mu)\left\lbrace  (1-p)-(\beta\Gamma)^2 \int Dz
\frac{ \int Dv [K
(z,v)
]^{m-1}\int D\xi[ \phi(z,v,\xi)]}{\int Dv [K
(z,v)
]^{m}}\right\rbrace 
\label{filling}
\end{equation}
with 
\begin{equation}
 \phi(z,v,\xi)=\frac{\cosh\sqrt{\Delta(z,v,\xi)}}{\Delta(z,v,\xi)}
- \frac{\sinh\sqrt{\Delta(z,v,\xi)}}{\Delta^{3/2}(z,v,\xi)}.
\label{filling2}
\end{equation}

The local stability of the RS solution for the FISG model is studied following close to de Almeida-Thouless analysis \cite{almeidalage}.
For this purpose, the stationary points (Eq. (\ref{stacionario})) are arbitrarily perturbed by 
independent quantities ($\xi_\alpha, \eta_{\alpha\gamma}$). The deviation $\Delta_{n}
(\xi_\alpha,\eta_{\alpha\gamma})=\Omega_{n}(p_\alpha,q_{\alpha\gamma})-\Omega_{n}(p_\alpha+2\xi_\alpha,
q_{\alpha\gamma}+\eta_{\alpha\gamma})$ can be obtained expanding $\Delta_n$ up to the second order in ($\xi_\alpha, \eta_{\alpha\gamma}$): 
\begin{equation}
\Delta_n = \sum_{\alpha,\gamma}G_{\alpha,\gamma}\xi_{\alpha}\xi_{\gamma}
+2\sum_{\alpha,(\gamma\nu)}G_{\alpha,\gamma\nu}\xi_{\alpha}\eta_{\gamma\nu}
+\sum_{(\alpha\beta),(\gamma\nu)}G_{\alpha\beta,\gamma\nu}\eta_{\alpha\beta}\eta_{\gamma\nu}
\label{diff}
\end{equation}
where
\begin{align}
G_{\alpha,\gamma}=\frac{\delta_{\alpha,\gamma}}{\beta^2 J^2} - 16 \langle S_{\alpha}^{z} S_{\alpha}^{z}
S_{\gamma}^{z}S_{\gamma}^{z}\rangle+16 \langle S_{\alpha}^{z} S_{\alpha}^{z}\rangle\langle
S_{\gamma}^{z}S_{\gamma}^{z}\rangle = (A-B)\delta_{\alpha,\gamma} +B
\\
G_{\alpha,\gamma\nu}= - 16 \langle S_{\alpha}^{z} S_{\alpha}^{z}
S_{\gamma}^{z}S_{\nu}^{z}\rangle+16 \langle S_{\alpha}^{z} S_{\alpha}^{z}\rangle\langle 
S_{\gamma}^{z}S_{\nu}^{z}\rangle= \left\lbrace \begin{tabular}{l} $C \mbox{ if $\alpha=\gamma$ or $\nu$ }$
 \\ $D \mbox{ if $\alpha\neq\gamma$ and $\nu$}$
\end{tabular} \right.
\end{align}
\begin{equation}
G_{\alpha\beta,\gamma\nu}=\frac{\delta_{\alpha\beta,\gamma\nu}}{2\beta^2 J^2} - 
16 \langle S_{\alpha}^{z} S_{\beta}^{z}
S_{\gamma}^{z}S_{\nu}^{z}\rangle+16 \langle S_{\alpha}^{z} S_{\beta}^{z}\rangle\langle
S_{\gamma}^{z}S_{\nu}^{z}\rangle
= \left\lbrace \begin{tabular}{l} $P \mbox{ if }\alpha\beta=\gamma\nu$
 \\ $Q \mbox{ if $\alpha=\gamma$}(\beta\neq\nu)$\\
$R.$
\end{tabular} \right. 
\end{equation}

The eigenvalues associated with the 
Hessian matrix (\ref{diff}) are known \cite{almeida}. In the limit $n\rightarrow 0$, there are three 
distinct eigenvalues:
\begin{equation}
\lambda_{\pm}=[A-B+(P-4Q+3R)\pm \sqrt{U}]/2
\label{lambda}
\end{equation}
and
\begin{equation}
\lambda_{AT}=P-2Q+R
\label{at}
\end{equation}
where
\begin{equation}
U=[(A-B)-(P-4Q+3R)]^2-8(C-D)^2,
\end{equation}
\begin{equation}
\begin{split}
 P-4Q+3R=&\frac{1}{2\beta^{2} J^{2}}-\int D z  
 \left[\varphi_{2}^2(z)-4\varphi_{2}(z)\varphi_{1}^2(z)+3\varphi_{1}^4(z)
 \right]
 \end{split},
\end{equation}
\begin{equation}
\begin{split}
 P-2Q+R=&\frac{1}{2\beta^{2} J^{2}}-\int D z  
 \left[\varphi_{2}(z)-\varphi_{1}^{2}(z)
 \right]^2
 \end{split},
\end{equation}
\begin{equation}
A-B =\frac{1}{\beta ^{2} J^{2}} - \int D z \left[ \varphi_{4}(z)- \varphi_{2}^2(z)\right],
\end{equation}
\begin{equation}
C-D =\int D z \left[ \varphi_{2}(z) \varphi_{1}^2(z) - \varphi_{3}(z)\varphi_{1}(z)\right],
\end{equation}
with
\begin{equation}
\varphi_{n}(z)= \left[{\int D\xi \frac{\partial^n}{\partial \bar{h}^n}\left( \cosh\sqrt{\bar{\Delta}(z,\xi)}\right) }\right]/{\bar{K}(z)}, 
\label{eqphi}
\end{equation}
$\bar{\Delta}(z,\xi)=\bar{h}^2(z,\xi)+\beta^2\Gamma^2$ with $\bar{ h}(z,\xi)=\beta J\sqrt{2}(\sqrt{q}z+\sqrt{p-q}\xi)$, and
\begin{equation}
\bar{K}(z)= \frac{s-2}{2} \cosh{\beta\mu}+ \int D\xi \cosh\sqrt{\bar{\Delta}(z,\xi)}.
\end{equation}
In that case, the stable RS solution would occur when the Hessian eigenvalues  $\lambda_{\pm}$ 
and $\lambda_{AT}$ 
are all positive.  As one can see, different from the classical SK model, in the FISG  model, the replica diagonal elements have an important role in the stability of the RS solution. Particularly, the condition for all eigenvalues of the Hessian matrix to be nonnegative in the paramagnetic solution ($q=0$) is used to locate the tricritical temperature $T_{tc}(\Gamma)$ at (for detail, see Ref. \cite{magal5}):
\begin{equation}
 T_{tc}/J=\frac{\sqrt{2}}{3}\varphi_{4}/\varphi_{2} 
\label{eetric}
\end{equation}
where $\varphi_{2}$  and $\varphi_{4}$ are defined in Eq. (\ref{eqphi}) when $q=0$ and $p=T_{tc}/J\sqrt{2}$.

\section{Results}
\label{results}
The numerical solutions of the saddle point order parameters equations allow 
one 
to build phase diagrams
($T/J$) ($T$ is the temperature) and 
$\Gamma/J$ 
(see, for instance, Refs. \cite{Alba2002,zimmermagal}).  In the $4S$ 
model,  it
is also taken into account 
$\mu/J$ to build phase diagrams (see, for instance, Ref. \cite{magal5}).
The parameter  $J$ is related to the variance of the Gaussian random coupling $J_{ij}$ given in Eq. (\ref{distrib}). 
Particularly, for the $4S$ 
case, the first order boundary has been located using the procedure 
introduced in Ref. \cite{franciscosalinas} for the GS model.
The Hessian eigenvalues (Eqs. (\ref{lambda}-\ref{at})) are inserted inside
these phase diagrams.

\begin{figure}[ht]
\begin{center}
    \includegraphics[angle=270,width=12cm]{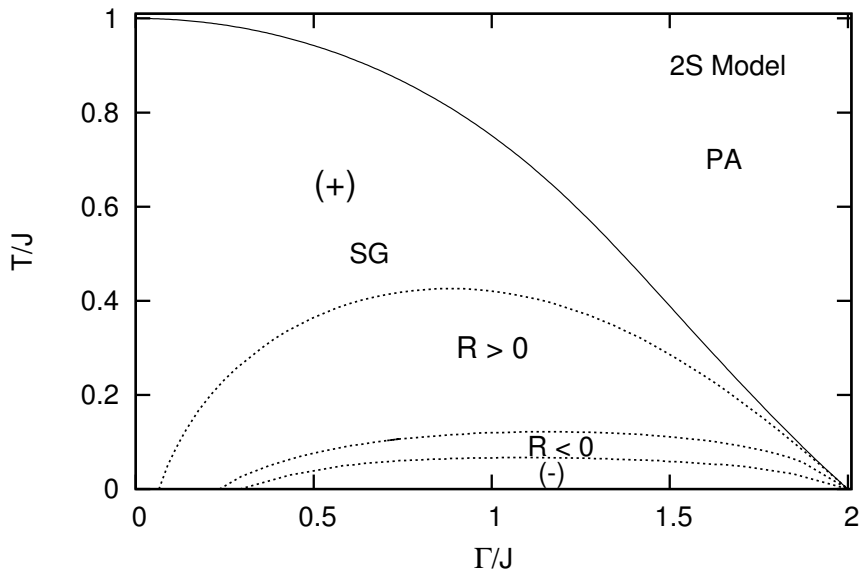}
\end{center}
\caption{Phase diagram $T/J$ {\it versus} $\Gamma/J$  for the 2S model showing the behavior of the eigenvalues $\lambda_{+}$ and $\lambda_{-}$. The notation $(+)$ $((-))$ indicates a region where both eigenvalues $\lambda_{+}$ and $\lambda_{-}$ are real 
positive (negative). In the region indicated by $R>0$ ($R<0$), these 
eigenvalues  
 become complex with their real part positive
(negative). The solid line indicates a second order transition.
}
\label{fig2s}
\end{figure}

Fig. \ref{fig2s} shows a phase diagram $T/J$ {\it versus} $\Gamma/J$ for the 2S model. 
This is equivalent to study the problem in the spin space, as discussed previously. The SG phase boundary is followed by Almeida-Thouless line (defined as $\lambda_{AT}=0$)  which
means that $\lambda_{AT}$ is negative in the whole SG phase \cite{Alba2002}.
For very small $\Gamma/J$, eigenvalues $\lambda_{+}$ and $\lambda_{-}$ (denoted by $\lambda_{\pm}$) are real positive ( indicated by ($+$) in Fig. \ref{fig2s}).
When $\Gamma$ is enhanced, the 
 contribution from the replica diagonal matrix elements for the SG solution 
 starts to increase.
As a consequence, 
eigenvalues
$\lambda_{\pm}$
become
complex conjugated pairs.
In regions indicated by $R$ (see Fig. \ref{fig2s}), $\lambda_{\pm}$ assume complex values with 
their
real part positive ($R>0$)
or negative ($R<0$). 
At very low temperature,
$\lambda_{\pm}$
 are real negative (indicate by ($-$) in  Fig. \ref{fig2s}). 
The  onset of regions  with
$R>0$ and $R<0$
 always occurs below $\lambda_{AT}=0$. 
However,  when $\Gamma$ increases, 
boundaries of such regions,   as well as 
that one with $(-)$,
become increasingly close to the $AT$ 
line.
In fact, 
these
boundaries
go
toward 
 the $QCP$ given by $\Gamma_{c}=2J$,  
which  
suggest
that the positivity of $\lambda_{AT}$ 
in the 
FISG
model  could not be enough to guarantee the replica symmetry stability near the $QCP$. 

\begin{figure}[ht]
\begin{center}
    \includegraphics[angle=270,width=12.5cm]{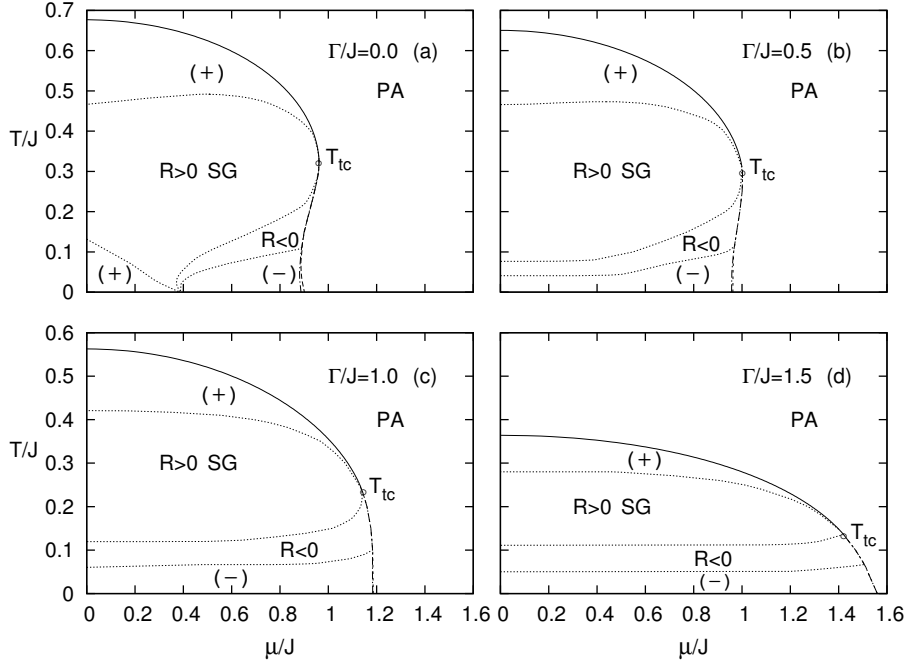}
\end{center}
\caption{ Phase diagrams T/J {\it versus} $\Gamma/J$ showing the behavior of eigenvalues $\lambda_{+}$ and $\lambda_{-}$ for the $4S$ model and several values of $\mu/J$.  
It is used the same convention as Fig. \ref{fig2s} for the instability regions of the RS solution.
The solid lines indicate second order phase transition. Dashed lines 
indicate a first order phase transition, in which results within RS and $1S\mbox{-}RSB$ solution are compared. $T_{tc}$  indicates the tricritical point. 
}
\label{figdiag}
\end{figure}

\begin{figure}[th]
\begin{center}
    \includegraphics[angle=270,width=12.5cm]{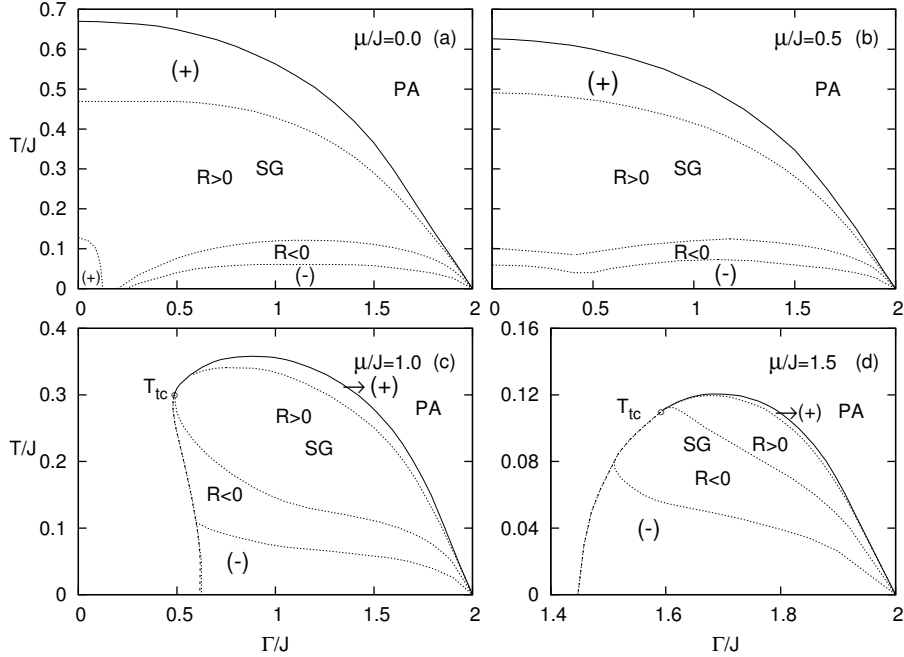}
\end{center}
\caption{Phase diagrams $T/J$ {\it versus} $\mu/J$ for the 4S model with
for several values of $\Gamma/J$. The same convention as Fig. \ref{fig2s} is used for the instability regions of the RS solution.
The solid lines indicate second order transition while dashed lines indicate a first order transition, in which the RS and $1S\mbox{-}RSB$ solutions are compared.
}
\label{fig4s}
\end{figure}

\begin{figure}[ht]
\begin{center}
    \includegraphics[angle=270,width=12.cm]{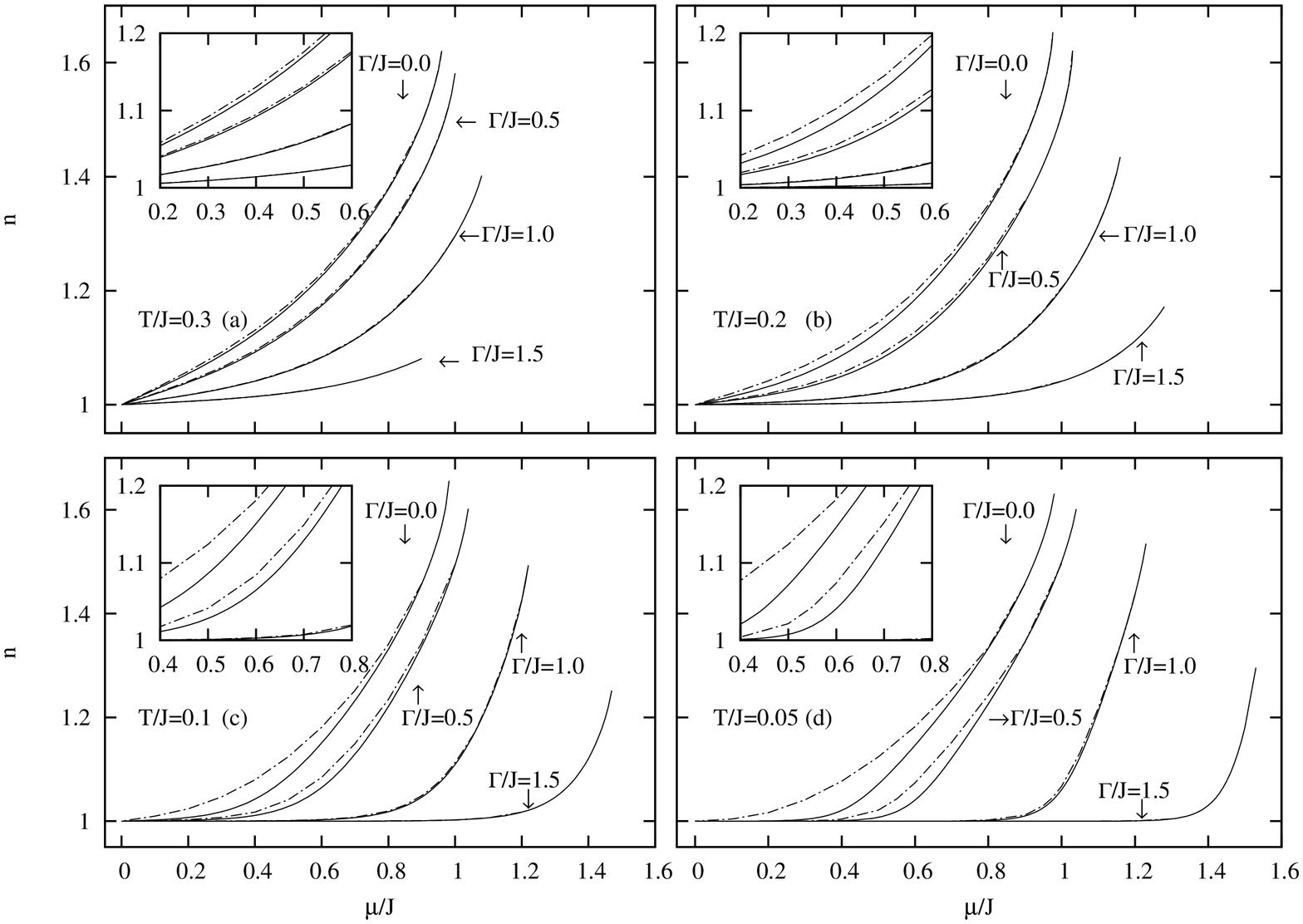}
\end{center}
\caption{ Average occupation $n$ versus $\mu/J$ for several values of $T/J$ and $\Gamma/J$ of the SG solution. The dashed-dotted lines indicate the $1S\mbox{-}RSB$ solution. 
The insets show in detail the difference between the value of $n$  in RS and $1S\mbox{-}RSB$ solution.
}
\label{figfat}
\end{figure}

 In the $4S$ model, the 
combined effects of $\mu$
and $\Gamma$ on $\lambda_{AT}$ and the $\lambda_{\pm}$ can be seen in Fig. \ref{figdiag}.
 This figure shows
phase diagrams $T/J$ {\it versus} $\Gamma/J$ for several values of $\mu/J$. 
For all values chosen for $\mu/J$,  
the freezing temperature $T_f $ decreases toward a $QCP$ when 
$\Gamma \rightarrow \Gamma_{c}$,
 as in the $2S$-model.
The half-filling situation 
($\mu=0$) is exhibited in Fig. \ref{figdiag}(a). 
In this case, the SG solution is always unstable against the replicon stability analysis. 
 However, in contrast with the 2S-model, it now appear a region with $R>0$ 
even for $\Gamma=0$.  Fig. 2(b) presents the same characteristics found in Fig. 2(a), except there is no region with $(+)$ at low $\Gamma/J$ and $T/J$. 
Nevertheless, for higher values of $\mu/J$, as shown in Figs. 2(c)-(d), the consequences for SG instability from the interplay between charge and spin correlations become more important. 
 For instance, when $\mu/J=1$ (see Fig. \ref{figdiag}(c)), there are no 
SG
solutions for 
$0\leq\Gamma \lesssim 0.5/J$. 
 When
$\Gamma 
\gtrsim 
0.5J$, with the arising of these solutions,
there is also a change in the nature of the SG
boundary phase with the onset of a tricritical point at temperature $T_{tc}$.  Below $T_{tc}$, a first order boundary is located in 
RS and $1S\mbox{-}RSB$ schemes, 
but, without any important difference between them \cite{magal5}.
 Regarding the behavior of $\lambda_{AT}$  at high values of $\mu/J$, it is found that the replicon remains negative in the entire SG phase, as previously observed in Figs. 2(a)-(b). 
However,  
 there is an enlargement of   
regions 
with $R>0$, $R<0$ and $(-)$. 
Particularly, 
 close to the first order boundary phase, there are only regions with $R<0$ or $(-)$.
 On the contrary, close to the second order boundary phase,  the region with $(+)$ becomes 
smaller.
 Finally, in Fig. 
\ref{figdiag}(d), 
 the SG phase boundary 
is strongly decreased as compared with Fig. \ref{figdiag}(c). The important point in this last case is that
the regions
with $R<0$ or $(-)$  
 occupy
almost the whole SG 
phase.

The combined effects of $\Gamma$ and $\mu$ can also be seen in a different perspective. 
In Fig. \ref{fig4s}, the phase diagrams are built now 
in $T/J$-$\mu/J$ surfaces for several values of $\Gamma/J$.
Fig. \ref{fig4s}(a) shows a phase diagram 
for $\Gamma=0$ with regions $R>0$, $R<0$ and $(-)$. 
 This result can be compared with Ref. \cite{franciscosalinas} using the mapping introduced in Ref. \cite{feldmann}. 
The 
phase boundary transition 
at $T_f$ 
is continuous at low $\mu/J$.
When $\mu/J$ increases, 
 a tricritical point appears at $T_{tc}$
and, at lower temperature, the first order part of $T_{f}$ 
presents a reentrance 
(see discussion in Ref. \cite{magal5}).
 Close to first order boundary, 
besides the usual replicon instability, the eigenvalues $\lambda_{\pm}$  become complex
with $R<0$ as  shown in Fig. \ref{figdiag}. 
The  enhancement 
of $\Gamma$, shown in Figs. \ref{fig4s}(b)-(d), decreases the second 
order boundary phase and the tricritical point as well as 
it suppresses
the reentrance in the first order one \cite{magal5}. 
 The behavior of $\lambda_{AT}$ remains basically the same found in Fig. \ref{figdiag}.
The SG phase is always replica symmetry breaking. 
 However, again, 
 the region with $(+)$ decreases, while 
 those ones
with $R>0$, $R<0$ and $(-)$ increase.
It should be emphasized 
that, 
close to the first order transition,  it is observed, again, a region where,  at same time, the replicon and 
the $\lambda_{\pm}$ violate the replica stability condition.

The complicated connection between charge and spin correlations due to the presence of $\Gamma$ can also be seen in the average occupation of fermions per site $n$
(see Eqs. (\ref{filling})-(\ref{filling2})).
Figure \ref{figfat} shows the behavior of 
$n$ 
within RS and $1S\mbox{-}RSB$ schemes
as a function of 
$\mu/J$ 
for several 
isotherms
and  $\Gamma/J$. 
For $\Gamma=0$, when 
temperature 
is decreased, $n$ 
remains at the half-filling 
even for $\mu/J\neq 0$.
This is consistent 
with an earlier result  \cite{zimmermagal,oppermann}, which suggests that the ocuppation of nonmagnetic states is exponentially small when $T$ is decreased. 
However, it is also true that this effect is 
weakened when  $1S\mbox{-}RSB$ solution is used, particularly, at lower $T$ \cite{feldmann}.
Nevertheless, the presence of $\Gamma$ 
enforces the tendency to preserve
the half-filling occupation,
as can be seen,
for higher $T$  
in Fig. {\ref{figfat}}(a).
Indeed, this effect is enhanced when temperature is 
lowered as shown in Figs. \ref{figfat}(b)-(d). 
Most important, 
when
$\Gamma$ is turned on,
the differences between RS and $1S\mbox{-}RSB$ solutions 
become
irrelevant. 
In other words, 
$\Gamma$ enforces the effect of temperature   
preserving $n$ at the half-filling 
for both 
RS and $1S\mbox{-}RSB$ levels of description.  

\section{Conclusions}
\label{conclusions}

In this work,  it has been studied the stability of spin-glass phase in the FISG model when a magnetic transverse field $\Gamma$ is applied and, particularly, 
$\lambda_{\pm}$ (the longitudinal eigenvalues of the Hessian matrix). The reason is that the presence of $\Gamma$ produces important effects in 
the diagonal replica matrix elements
which are closely related to the behavior of $\lambda_{\pm}$.
 In fact,
the FISG model 
 can be presented 
in two formulations, the so-called $4S$ and $2S$ models \cite{albagusmao,zimmermagal}.  
 In the $4S$ model, 
the original four states of the Fock space are maintained. 
 In the $2S$ one, 
the nonmagnetic states are not allowed.
Particularly, this means that in the $4S$ model there is
a connection between charge and spin degrees of freedom \cite{feldmann,oppermann2}.

The results presented in Figs. \ref{fig2s}-\ref{figdiag} display two distinct situations for $\lambda_{AT}$ and 
$\lambda_{\pm}$. 
For the $2S$ and $4S$ models, 
the results show 
that the replicon $\lambda_{AT}$ is always 
real 
negative in the entire  SG phase 
for any value of $\Gamma/J$. 
On the other hand,  for the $2S$ model (see Fig. \ref{fig2s}),
the enhancement of $\Gamma/J$ produces regions where $\lambda_{\pm}$ becomes complex with real part positive ($R>0$), negative ($R<0$) or
even 
real 
negative ($(-)$). 
Furthermore, all boundary lines of such regions converge to the $QCP$ 
when $\Gamma\rightarrow \Gamma_{c}$. 
 In the $4S$ model (see Fig. \ref{figdiag}),  $\lambda_{\pm}$ also present regions with $R>0$, $R<0$ and $(-)$.
 In fact, there are other similarities between the two models.  When $\Gamma\rightarrow \Gamma_{c}$,
the boundary lines of these regions 
have the same behavior already found in the $2S$ model.
 Most important,
this result is independent 
of a particular value of $\mu/J$. 
 Nevertheless,  for 
$\Gamma<\Gamma_{c}$, when $0\leq\mu <J$, 
the regions with  
$R<0$ and $(-)$
are relatively small and they appear only at lower temperature. 
By contrast, 
the situation is different  when $\mu\gtrsim J$. 
 Particularly, 
regions with $R<0$ and $(-)$ 
are considerably enlarged while that one with $(+)$
 is decreased.
It should be remarked that close to the first order boundary phase, besides the 
region with $(-)$, 
 there is 
only region with $R<0$. 

The role of $\Gamma$
described
in the previous paragraph for the $4S$ model 
is confirmed 
in Fig. \ref{fig4s}. 
There,  the increase of $\mu$ produces regions with $R>0$, $R<0$ and $(-)$ as expected, for instance, from the mapping with the GS model when $\Gamma=0$ \cite{feldmann}.
However, the increase of $\Gamma$ tends to depress $T_{f}$, the tricritical point $T_{tc}$ and to destroy the reentrance in the first boundary phase \cite{magal5}. 
 Such increase
also 
depresses 
boundary lines of regions 
 with
$R>0$, $R<0$ and $(-)$. 
The important point is that these boundary lines follow the change in location of the $T_{tc}$.  
Therefore, no matter the value of $\mu$, 
 the increase of $\Gamma$ 
tends to reproduce the scenario in which all boundary lines of regions  $R>0$, $R<0$ and $(-)$ are decreasing towards zero temperature.

The behavior of $n$ exhibited in Fig. \ref{figfat} can clarify the role of $\Gamma$ in the problem  as source of similarities between the $2S$ and $4S$ model when $\Gamma\rightarrow \Gamma_{c}$. 
The increase of $\Gamma$ leads $n$ to remain at the half-filling 
even when $\mu$ is increased from zero. 
Therefore, the increase of $\Gamma$ acts to avoid the double occupation. 
In the opposite limit, when $\Gamma<<\Gamma_{c}$, the increase of chemical pressure 
tends to  remove $n$ from the half-filling
which  leads 
the $4S$ model to become closer to the classical $GS$ model.

Therefore, it is possible to identify two regimes. In the first one, which  is common to both $2S$ and $4S$ models, 
the region with $(+)$ 
is increasingly small when $\Gamma\rightarrow \Gamma_{c}$. 
In the second regime, when $\Gamma$ is 
not close to $\Gamma_{c}$, there is a mixing of effects due to the presence of $\mu$
and $\Gamma$. 
In that case, 
the increase of $\mu$ produces 
 the increase of regions where the eigenvalues $\lambda_{\pm}$ 
are
complex or negative.  It should be remarked that the regions with $R<0$ and (-) 
are exactly those ones which
occupy
most part of the SG phase diagram, particularly, close to the first order boundary phase. Therefore, even as suggested in Refs. \cite{oppermann2,yokoi}, that the condition $R>0$ would be enough to garantee the SG stability, the dominance of regions with $R<0$ and $(-)$ for certains values of $\Gamma$ and $\mu$ indicate clearly an additional instability to the SG solution which must be  taken into account simultaneously with that one related with the replicon.

Surely, the reliability of some results are quite limited by the use of $SA$, particularly, 
those ones 
close to the $QCP$. Nevertheless, 
in 
 the first regime, 
the results show 
clearly  that 
the tendency to decrease the region with $(+)$, when $\Gamma$ is increased, can be observed even at higher temperature.
This tendency  would indicate that close to the $QCP$ 
the analysis of $\lambda_{\pm}$ could also play a more important role.  However, there is no doubt 
that
 it is needed to 
go beyond the SA to obtain a more conclusive result close to the $QCP$.
On the other hand, for the second regime,  results are more conclusive.
The effects of $\Gamma$ combined with the increase of $\mu$ 
give an additional instability at higher temperatures. This new instability is important, 
particularly, close to the first order  boundary where the use of $SA$ is not so limited as in the case of a continous transition \cite{theo}.   

 To conclude, in this work, it has been studied in detail the behavior of $\lambda_{\pm}$ in the FISG model as a function of $\Gamma$. It has been presented several evidences indicating that these particular eigenvalues of the Hessian matrix can be also source of problem to the local stability of the $RS$ solution when $\Gamma$ increases. It is clear that this results are very much restricted to the FISG model. However, one can speculate whether or not others diferent formulations of infinite-range Ising spin-glass  model
in a transverse magnetic field could experiment similar additional instabilities likewise those ones found in the $2S$ model.

One last comment must be done. Although 
 there are 
strong similarities between the $2S$ and $4S$ models in the first regime,
it is not obvious  
that this similarity can be explained 
by 
the behavior of the
$n$ which remains at the half-filling when $\Gamma$ increases.
One alternative,  would consist of adding 
to the 
model given in Eq. (\ref{ham}) a local on site repulsive Coulomb interaction with strength $U$. Therefore, 
in the limit $U\rightarrow \infty$,  it would be possible to avoid locally the double occupation \cite{feldmann}. This 
approach
is currently under investigation.

\section*{Acknowledgments}

S. G. Magalh\~aes acknowlegdes the hospitality of CBPF (Centro Brasileiro de Pesquisas F\'{\i}sicas) where this work was concluded.  
The authors also acknowledge partial financial support from the Brazilian agencies CAPES (Coordena\c{c}\~ao de Aperfei\c{c}oamento de Pessoal de N\'{\i}vel Superior), FAPERGS (Funda\c{c}\~ao de Amparo \`a Pesquisa do Estado Rio Grande do Sul) and CNPq (Conselho Nacional de Desenvolvimento Cient\'{\i}fico e Tecnol\'ogico).

\end{document}